\documentclass{ws-ijmpa}
\def\Vec#1{\mbox{\boldmath $#1$}}
\def\eqne{\end{equation}}
 
\def\eqnb{\begin{equation}}
\def\PTP{Prog. Theor. Phys.(Kyoto)}

\def\PLB{{Phys. Lett.} B}
\def\PRL{Phys. Rev. Lett.}
\def\PRD{{Phys. Rev.} D}

\begin{document}

\markboth{Sadataka Furui}{ The Flavor Symmetry in the Standard Model and the Triality Symmetry}

\catchline{}{}{}{}{}

\title{ The Flavor Symmetry in the Standard Model \\and the Triality Symmetry
}

\author{\footnotesize SADATAKA FURUI
}

\address{ Faculty of Science and Engineering, Teikyo University.\\
1-1 Toyosatodai, Utsunomiya, 320-8551 Japan\\
furui@umb.teikyo-u.ac.jp}



\maketitle

\pub{Received 24 Aug.2012}{Accepted 21 Sept. 2012}

\begin{abstract}
A Dirac fermion is expressed by a 4 component spinor which is a combination of two quaternions 
and which can be treated as an octonion. The octonion possesses the triality symmetry, which defines symmetry of fermion spinors and bosonic vector fields. 
 
The triality symmetry relates three sets of spinors and two sets of vectors, which are transformed among themselves via transformations $G_{23}, G_{12}, G_{13}$, $G_{123}$ and  $G_{132}$. 
If the electromagnetic (EM) interaction is sensitive to the triality symmetry, i.e. EM probe selects one triality sector, EM signals from the 5 transformed world would not be detected, and be treated as the dark matter.
According to an astrophysical measurement, the ratio of the dark to ordinary matter in the universe as a whole is almost exactly 5.

We expect quarks are insensitive to the triality, and triality will appear as three times larger flavor degrees of freedom in the lattice simulation. 
\end{abstract}

\vskip 0.2 true cm
\footnotesize{  PACS numbers: 03.65.Fd, 11.30.Hv, 11.30.Rd, 12.60.Jv, 23.40.Bw, 11.30.Cp, \\ \qquad13.15,+g,13.30Ce,13.40.Ks, 14.80.Bn}


\section{Introduction}

The presence of dark matter in the universe and its abundance about 5 times more than that of normal matter\cite{SA11} is an important puzzle.  In the standard model, matter are described by spinors 
which are expressed by the algebraic number system quaternions. \'Elie Cartan\cite{Cartan66} showed that two quaternion makes an octonions which together with two sets of 4-dimensional vectors has three bases, which transform among themselves via 5 transformations 
\[
G_{12}, G_{13}, G_{23}, G_{132}={^tG}_{12}G_{13} {\quad \rm and\quad } G_{123}={^tG}_{13}G_{12}
\]
We express the Dirac fermion basis $\eta=\left(\begin{array}{c} A\\
                                                              B\end{array}\right)$ and 
$\chi=\left(\begin{array}{c} C\\
                                                              D\end{array}\right)$ and the
vector field $E$ and $E'$ as follows. 
\begin{eqnarray}
 A=\xi_{14}{\sigma_x}+\xi_{24}{\sigma_y}+\xi_{34}{\sigma_z}+ \xi_0 { I}\\
 B= \xi_{23}{\sigma_x }+\xi_{31}{\sigma_y}+\xi_{12}{\sigma_z}+\xi_{1234}{ I}\\
 C= \xi_1{\sigma_x}+\xi_2{\sigma_y}+\xi_3{\sigma_z}+\xi_4{ I}\\
 D=  \xi_{234}{\sigma_x}+\xi_{314}{\sigma_y}+\xi_{124}{\sigma_z}+\xi_{123}{ I} \\
 E=x_1{\Vec i}+x_2{\Vec j}+x_3{\Vec k}+x_4{ I}\\
 E'=x_1'{\Vec i}+x_2'{\Vec j}+x_3'{\Vec k}+x_4'{ I}
\end{eqnarray}

The transformations on the quaternion $G_{ij}$ and $G_{ijk}$ operated on these bases are 
\[
\begin{array}{ccccc}
G_{23}A\to C& G_{12}A\to E'(E)& G_{13}A\to A(B)& G_{123}A\to E' & G_{132}A\to C(D)\\
G_{23}B\to D& G_{12}B\to E(E')& G_{13}B\to B(A)& G_{123}B\to E & G_{132}B\to D(C)\\
G_{23}C\to A& G_{12}C\to C(D)& G_{13}C\to E'& G_{123}C\to A(B) & G_{132}C\to E'(E)\\
G_{23}D\to B& G_{12}D\to D(C)& G_{13}D\to  E& G_{123}D\to B(A) & G_{132}D\to E(E')\\
G_{23}E\to E(E') & G_{12}E\to B(A)& G_{13}E\to D& G_{123}E\to D(C) & G_{132}E\to B\\
G_{23}E'\to E'(E) & G_{12}E'\to A(B)& G_{13}E'\to C & G_{123}E'\to C(D) & G_{132}E'\to A\\
\end{array} 
\]
\vskip 0.3 true cm
Here, the notation $E'(E)$ or $A(B)$ means that the spacial components are transformed to the unbracketed set while the temporal component is transformed to the bracketed component.
The vector field that satisfies $E=E'$ is the self-dual field.

In this framework, vector fields are expressed as Pl\"ucker coordinate in the space of spinors, and there is no entanglement of the intrinsic coordinate and the spacial coordinate as in the minimal symmetric standard model ( MSSM ). The 16 spinor bases are categorized as
\[
\{\xi_0,\xi_1,\xi_2,\xi_3,\xi_4\}, \{\xi_{12},\xi_{31},\xi_{23},\xi_{24},\xi_{34}\},\{\xi_{123},\xi_{124},\xi_{314},\xi_{234},\xi_{1234}\}
\]
and the vector fields are expressed as
\[
\{x_1,x_2,x_3,x_4\} {\quad \rm and \quad } \{x_1',x_2',x_3',x_4'\}.
\] 

The transformation $G_{23}$ corresponds to the interchange of the upper component and the lower component of the Dirac fermion, or the transformation to the world of anti-particles. Other transformations, $G_{12}, G_{13}, G_{123}$ and $G_{132}$  could violate the Lorentz symmetry, but this transformation to a different triality sector would not be detected on the earth. 

We mainly observe universe via optical, or electro-magnetic detectors.  The spinors expressed by     quaternion allow the charge conjugation and the triality and the charge conjugation altogether allow  presence of 6 sectors, but the EM detector on the earth is sensitive to one sector and the 5 transformed world via  $G_{23},G_{12},G_{13}, G_{123}$ and $G_{132}$ would appear as the dark matter.

Pythagoras(ca.570b.c-ca.490b.c.) said numbers are the arkhe (i.e. the origin of mysteries of the cosmos)\cite{wiki}.
The Modern physics is expressed by complex numbers and non-commutative algebraic numbers, and whether the super-symmetry of \'E. Cartan can govern the universe is an interesting problem. 
 
Let us review the mathematical history on the establishment of the world of complex numbers and quatenions\cite{Number91}.  
Gauss showed in 1831 that a hypercomplex number system, which is called Complex number system ${\Vec C}$ is unique.
Hamilton discovered in 1843 a hypercomplex number which is called Quaternion $\Vec H$. It is not a field; i.e. commutative law of multiplication does not hold, and every non-zero element has their inverse.
 
The hyper complex system of numbers is called real algebras and if division can be performed unambiguously, one speaks of division algebra.  The uniqueness of a quaternion is proven by Frobenius,
and Hopf showed that every finite-dimensional commutative division algebra with unit elements other than $\Vec R$ is isomorphic to  $\Vec C$\cite{Number91}.

Via a duplication of the quaternion $\Vec H$, one can construct octonion $\Vec O$.
The uniqueness theorem of octonion is discovered by Zorn.  It possesses the triality symmetry, in which fermions and vector particles transform with each other\cite{Cartan66}. 

 I investigate, consequences of the super symmetry induced from the triality symmetry of octonions that fermions possess.  The structure of this paper is as follows. 
In Sect.2, I clarify the triality symmetry of leptons and quarks. In Sect.3, the trilaity symmetry of the gluon and its self-energy are estimated. Consequences in baryon decay, meson decay and neutrino mixing are summarized in Sect.4.
In Sect.5, the electromagnetic current and neutral current in the standard model is investigated, and the selection of triality in the electromagnetic interaction is discussed.
In Sect.6, possible interpretation of dark matter and the velocity of neutrino using the triality symmetry is prsented. Conlusion and discussion are given in Sect.6.

\section{The triality symmetry of leptons and quarks}
Let us consider perturbative correction on the self-energy of a gluon in the world in which instantaneous gluon $x_4, x_4'$ exchange and the one loop correction is already considered and quarks possess experimentally defined mass which is given by the Lagrangian 
\begin{equation}
{\mathcal L}=m\bar\psi \psi
\end{equation}

The neutrino is left-handed and has a very small mass, and it is expected to be a Majorana particle; i.e. its antiparticle is identical to itself.  In order to give neutrino a mass, presence of a right-handed heavy neutrino $\psi^c_R$ was assumed to introduce the Lagrangian
\begin{eqnarray}
{\mathcal L}^L&=&\frac{1}{2}m_L(\bar\psi_L \psi^c_R+\bar\psi^c_R\psi_L)\\
 {\mathcal L}^R&=&\frac{1}{2}m_R(\bar\psi^c_L \psi_R+\bar\psi_R\psi^c_L)
\end{eqnarray}
The process of producing phenomenological neutrino mass and explaining heavy right-handed neutrino as the Dark matter is called seesaw mechanism\cite{Zuber12}.

Leptons or quarks has the triality symmetry\cite{SF11a}, and although a neutrino is the same as its antiparticle, it has two neutrino partners in different triality sectors. The electron, muon and tau are
expected to be eigenstates of mixtures of three leptons in different triality sectors. 

The Yukawa mass is defined via Lagrangian defined by the Higgs field $\phi$, 
\[
\phi=\frac{1}{\sqrt 2}\left(\begin{array}{c} 0\\
                                                          v+H(x)\end{array}\right)
\]
as ${\mathcal L}=-c_\nu \bar \nu^C_L v \nu_L$. Here $H(x)=\left(\begin{array}{c} H^+\\
                       H^0\end{array}\right)$, and $v, H^+, H^0$ are complex scalar fields.

Lagrangians coupling neutrunos in different triality sectors
\begin{eqnarray}
{\mathcal L_1}&=&\frac{1}{2} v_1(\bar\psi_{t2} \psi_{t3}+\bar\psi_{t3}\psi_{t2})\\
{\mathcal L_2}&=&\frac{1}{2} v_2(\bar\psi_{t1} \psi_{t3}+\bar\psi_{t3}\psi_{t1})\\
{\mathcal L_3}&=&\frac{1}{2} v_3(\bar\psi_{t1} \psi_{t2}+\bar\psi_{t2}\psi_{t1})
\end{eqnarray}
could produce two degenerate light neutrino and a heavy neutrino via diagonalization of
\[\left(\begin{array}{ccc} 0& v_3 & v_2\\
                          v_3 & 0 & v_1\\
                          v_2 & v_1 & 0\end{array}\right).
\]

Since it is difficult to explain masses of leptons, I consider masses of the gluons whose mass is expected to be given by the quark loops.  In order to make the matrix elements Lorentz invariant, I adopt the dotted spinors \`a la van der W\"arden\cite{LL72, vdW29}.
When quarks in a triality sector $\psi$ obey a unitary transformation $U$, the quarks $\bar \psi$ obey the complex conjugate of the transformation denoted as $U^*$.

We define the four component Dirac fermion representation of a quark or a lepton as\cite{labelle10} 
\[
\Psi_D=\left(\begin{array}{c}\eta_e\\
                                   \chi_e\end{array}\right)
\]
and the free Lagrangian 
\begin{eqnarray}
\mathcal L_{Dirac}&=&\bar\Psi_D(\gamma^\mu \partial_\mu-m)\Psi_D\nonumber\\
&=&\chi_e^\dagger\bar\sigma^\mu i\partial_\mu\chi_e +\eta_e^\dagger \sigma^\mu i\partial_\mu\eta_e-m(\eta_e^\dagger\chi_e+\chi_e^\dagger\eta_e)
\end{eqnarray}

Whether neutrino is also expressed by Dirac fermion is controversial. When an anti-particle is identical with its particle, it is called Majorana particle. If lepton number violating double $\beta$ decay of a nucleus is observed, one can conclude that the neutrino is a Majorana particle, but the experiments
are not conclusive \cite{PDG10, DHKK12,ACO11,Faessler12}. 
The Majorana neutrino is represented by a Weyl spinor $\chi_p$ whose antiparticle is identical to the particle, or by a 4-component combination
\[
\Psi_M=\left(\begin{array}{c} i\sigma^2{\chi_p}^{\dagger T}\\
                                   \chi_p\end{array}\right)
\]
and
\[
{\mathcal L}_M=\chi_p^\dagger\bar \sigma^\mu i\partial_\mu\chi_p-\frac{m}{2}(\chi_p\cdot \chi_p+\bar\chi_p\cdot\bar\chi_p).
\]
Here the charge conjugate state is defined as
\[
\Psi_p^c=\left(\begin{array}{c}i\sigma^2\chi_p^{\dagger T}\\
                                          -i\sigma^2\eta_p^{\dagger T}\end{array}\right)=\left(\begin{array}{c}\eta_{\bar p}\\
                                                                                                               \chi_{\bar p}\end{array}\right)
\]
since $i\sigma^2\chi^{\dagger T}$ behaves as right chiral spinor.

The dot product is defined as $\chi\cdot\chi=\chi_2\chi_1-\chi_1\chi_2$
\[
\eta\cdot\chi=\eta^a\chi_a=\epsilon^{ab}\eta_b\chi_a=-\epsilon^{ab}\chi_a\eta_b=\chi^b\eta_b=\chi\cdot\eta
\]
The $\epsilon$ is defined as $(i\sigma^2)^{ab}=\epsilon^{ab}$, and $\sigma^2\sigma^\mu\sigma^2=\bar\sigma^{\mu T}$.

The Lagrangian of the Weyl fermion is identical to that of Majorana fermion, but it can be interpreted as a self-dual field.

The Dirac particle is expressed by a four component spinor, or a combination of two Pauli spinors or two algebraic objects quaternions. A combination of two quaternions makes an octonion and the octonion has the triality symmetry\cite{Cartan66,SF11a}.

\section{The triality symmetry of gluons}
In \cite{SF11a}, I showed that the lattice QCD simulation in the momentum subtraction scheme(MOM) using 2+1flavor domain wall fermion(DWF) gauge configuration after Coulomb gauge fixing showed the effective coupling $\alpha_s(q)$ similar to that of ADS/QCD scheme and the $\alpha_{g1}(q)$ obtained from the electron scattering at JLab\cite{BdTD10}. 

A trilinear form of the Lagrangian defined by Cartan fixes the gluon $x_1$ couples with a pair of spinors $\xi_{12}\xi_{314}$ etc. and when a self-dual gluon is exchanged in the pair of spinors, the pair cannot be coupled with the original $x_1$. Thus, one self-dual gluon exchange one-loop gluon self-energy does not appear and the two self-dual gluon exchange three loop gluon self-energy were considered\cite{SF11a}.  
 I ignore exchange of $x_4, x'_4$ and diagrams of exchanging two same type of gluons $x_3,x_3$, $x_2,x_2$ or $x_1,x_1$.

\begin{figure}[th]
\centerline{\psfig{file=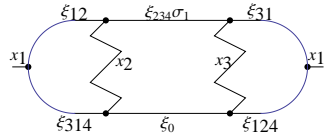,width=4.3cm}}
\caption{Transverse gluon self energy of ${^t(} B,D)\to {^t(} D,A)$}
\label{g1123_14}
\end{figure}

\begin{figure}[th]
\centerline{\psfig{file=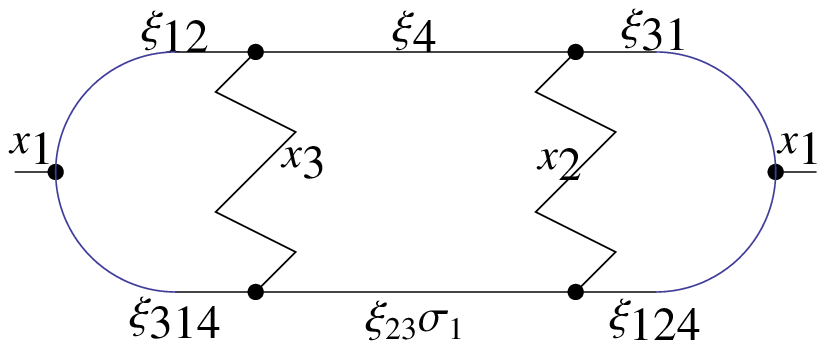,width=4.3cm}}
\caption{Transverse gluon self energy of ${^t(} B,D)\to {^t(} C,B)$}
\label{g1132_41}
\end{figure}

\begin{figure}[th]
\centerline{\psfig{file=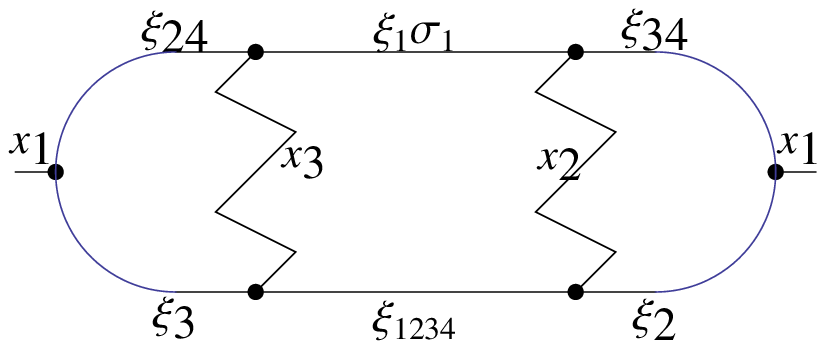,width=4.3cm}}
\caption{Transverse gluon self energy of ${^t(} A,C)\to {^t(} C,B)$}
\label{g1132_14}
\end{figure}

\begin{figure}[th]
\centerline{\psfig{file=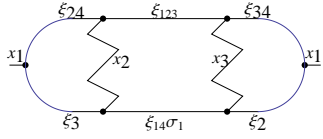,width=4.3cm}}
\caption{Transverse gluon self energy of ${^t(} A,C)\to {^t(} D,A)$}
\label{g1123_41}
\end{figure}

 A transverse gluon polarized along the $x$-axis can have intermediate $^t|\xi_{234}\xi_0>$  or $^t|\xi_4\xi_{23}\rangle$ as shown in Fig. \ref{g1123_14} and \ref{g1132_41} and in Fig. \ref{g1132_14} and \ref{g1123_41}.
 The  basis of the intermediate state $^t|\xi_{234},\xi_0\rangle$ is 
\[
\left(\begin{array}{cc} \sigma_x n_x\sinh\frac{\phi}{2}& 0\\
                               0 & \cosh\frac{\phi}{2}
\end{array}\right) \left|\begin{array}{c}\xi_{234}\\
                     \xi_0\end{array}\right\rangle =U\left |\begin{array}{c}\xi_{234}\\
                     \xi_0\end{array}\right\rangle 
\] 
and that for the state $^t|\xi_{4},\xi_{23}\rangle$ is 
\[
 \left(\begin{array}{cc} \cosh\frac{\phi}{2}& 0\\
                               0 & \sigma_x(- n_x)\sinh\frac{\phi}{2}\end{array}\right)\left |\begin{array}{c} \xi_4\\
                          \xi_{23}\end{array}\right\rangle = V\left|\begin{array}{c} \xi_4\\
                          \xi_{23}\end{array}\right\rangle
\]
\noindent Their conjugate bases are
\[
\left\langle\begin{array}{c} \xi_{234}\\
                          \xi_{0}\end{array}\right| \left (\begin{array}{cc} \sigma_x(-n_x)\sinh\frac{\phi^*}{2}& 0\\
                               0 & \cosh\frac{\phi^*}{2}\end{array}\right)= \left\langle\begin{array}{c} \xi_{234}\\
                          \xi_{0}\end{array}\right| U^*
\]
\noindent and
\[
\left\langle\begin{array}{c} \xi_4\\
                          \xi_{23}\end{array}\right| \left (\begin{array}{cc} \cosh \frac{\phi^*}{2}& 0\\
                               0 & \sigma_x n_x\sinh\frac{\phi^*}{2}\end{array}\right)= \left\langle\begin{array}{c} \xi_4\\
                          \xi_{23}\end{array}\right| V^*
\]

There are diagrams with intermediate state $^t|\xi_{123}\xi_{14}\rangle$ and $^t|\xi_1\xi_{1234}>$.
The transition probability from $^t|\xi_{12}\xi_{314}\rangle$ to $^t\langle \xi_{31}\xi_{124}|$, and $^t|\xi_{24}\xi_{3}\rangle$ to $^t\langle \xi_{34}\xi_{2}|$ which have different intermediate states 
would both be $\cosh^2\frac{\phi}{2}-\sigma_x^2\sinh^2\frac{\phi}{2}=1$ when $\phi$ is real.

The basis of the $^t|\xi_1\xi_{1234}>$ is
\[
\left(\begin{array}{cc} \sigma_x n_x\sinh\frac{\phi}{2}& 0\\
                               0 & \cosh\frac{\phi}{2}
\end{array}\right) \left|\begin{array}{c}\xi_{1}\\
                     \xi_{1234}\end{array}\right\rangle =V\left|\begin{array}{c}\xi_{1}\\
                     \xi_{1234}\end{array}\right\rangle 
\]
\noindent and the basis of the $^t|\xi_{123}\xi_{14}>$
\[
\left(\begin{array}{cc} \cosh\frac{\phi}{2}& 0\\
                               0 & \sigma _x(- n_x)\sinh\frac{\phi}{2}
\end{array}\right) \left|\begin{array}{c}\xi_{123}\\
                     \xi_{14}\end{array}\right\rangle =W\left|\begin{array}{c}\xi_{123}\\
                     \xi_{14}\end{array}\right\rangle. 
\] 
\noindent Their conjugate basis are
\[
\left\langle\begin{array}{c} \xi_1\\
                          \xi_{1234}\end{array}\right| \left (\begin{array}{cc} \sigma_x n_x \sinh\frac{\phi^*}{2}& 0\\
                               0 & \cosh\frac{\phi^*}{2}\end{array}\right)= \left\langle\begin{array}{c} \xi_1\\
                          \xi_{1234}\end{array}\right| V^*
\]
\noindent and
\[
\left\langle\begin{array}{c} \xi_{123}\\
                          \xi_{14}\end{array}\right| \left (\begin{array}{cc} \cosh \frac{\phi^*}{2}& 0\\
                               0 & \sigma _x(- n_x)\sinh\frac{\phi^*}{2}\end{array}\right)= \left\langle\begin{array}{c} \xi_{123}\\
                          \xi_{14}\end{array}\right| W^*
\]

 In Fig. \ref{g1123_14}, the scalar component of the intermediate state $\xi_0$ is a component of $A$ and not that of $B$, as the $\xi_{12}$ and $\xi_{31}$ in the initial and final state.
This kind of mixing the scalar component is a manifestation of the enlarged symmetry of the octonion. 
The state will not be detected as an asymptotic state, but could appear in the intermediate state.
The transformation of the basis in Fig.\ref{g1123_14} to Fig.\ref{g1123_41} are ${^t(}B, D)\to {^t(}D,A)\to {^t(}B,D)$, ${^t(}B, D)\to {^t(}C,B)\to {^t(}B,D)$, ${^t(}A, C)\to {^t(}C,B)\to {^t(}A,C)$, ${^t(}A, C)\to {^t(}D,A)\to {^t(}A,C)$, respectively.


\begin{figure}[th]
\centerline{\psfig{file=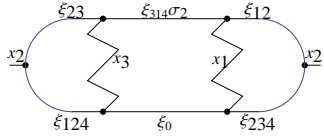,width=4.3cm}}
\caption{Transverse gluon self energy of ${^t(} B,D)\to {^t(} D,A)$}
\label{g2231_24}
\end{figure}

\begin{figure}[th]
\centerline{\psfig{file=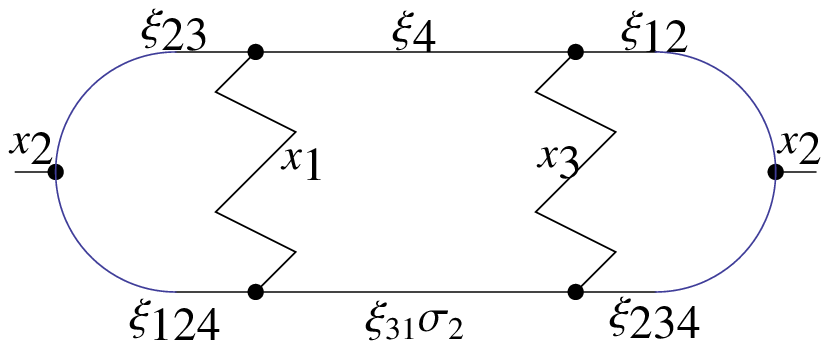,width=4.3cm}}
\caption{Transverse gluon self energy of ${^t(} B,D)\to {^t(} C,B)$}
\label{g2213_42}
\end{figure}

\begin{figure}[th]
\centerline{\psfig{file=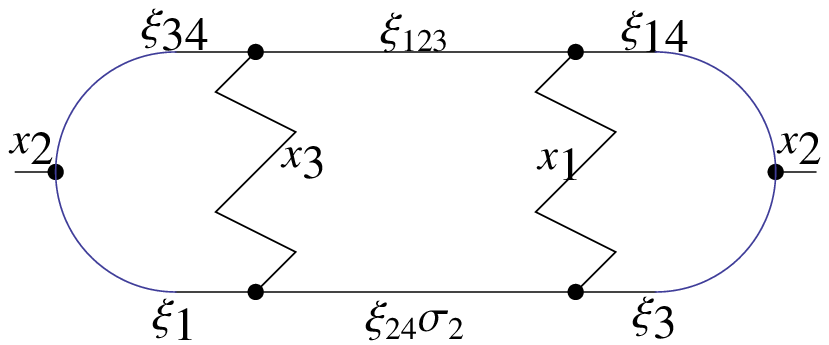,width=4.3cm}}
\caption{Transverse gluon self energy of ${^t(} A,C)\to {^t(} D,A)$}
\label{g2231_42}
\end{figure}

\begin{figure}[th]
\centerline{\psfig{file=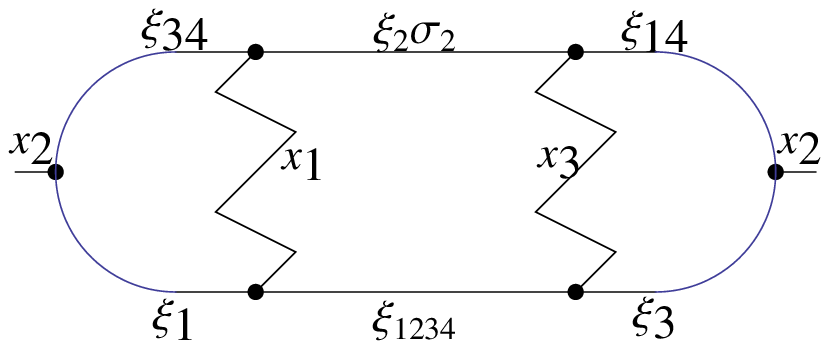,width=4.3cm}}
\caption{Transverse gluon self energy of ${^t(} A,C)\to {^t(} C,B)$}
\label{g2213_24}
\end{figure}

The corresponding diagrams for a gluon polarized along the $y$-axis are Fig.\ref{g2231_24} and \ref{g2213_42} and Fig.\ref{g2231_42} and \ref{g2213_24}.


In \cite{SF11b}, the self-energy of the transverse gluon  $\Pi^a_{11}$ was investigated via a current:
\begin{eqnarray}
&&{J_a( k_z p_y )}=m^4 g^6
\frac{ k_z\times  p_y}{m} Tr\gamma_1\gamma_5\frac{-\gamma_3  k_z+m}{k_z^2+m^2}\gamma_3\frac{1}{k_z^2}\frac{1}{m}\frac{1}{p_y^2}\gamma_2\frac{-\gamma_2 p_y+m }{p_y^2+m^2}\nonumber\\
&&\times \gamma_1\gamma_5 \frac{ k_z\times  p_y}{m} \frac{\gamma_3 k_z+m}{k_z^2+m^2}\gamma_2\gamma_5
\frac{\gamma_2 p_y+\gamma_3 k_z+m}{p_y^2+k_z^2+m^2}\gamma_3\gamma_5\frac{\gamma_2 p_y+m}{p_y^2+m^2}
\frac{ d {p_y} d {k_z} }{(2\pi)^{2}}.
\end{eqnarray}

By choosing $(k_z/m)^2=k^2, (p_y/m)^2=p^2$, I integrate numerically
\begin{eqnarray}
\Pi_{11}^a&=&4m^4  \int \int g^6\frac{m^6+3m^4(k_z^2+p_y^2)-m^2 k_z^2 p_y^2}{(k_z^2+m^2)^2(p_y^2+m^2)^2(k_z^2+p_y^2+m^2)} f(k^2,p^2) \frac{d{k_z} d{p_y}}{(2\pi)^2} \nonumber\\
&=&4\alpha_s \int \int \alpha_s(mk) \alpha_s(mp)\frac{1+3(k^2+p^2)-k^2p^2}{(k^2+1)^2(p^2+1)^2(k^2+p^2+1)}\frac{f(k^2,p^2)dk dp (4\pi)^3}{(2\pi)^2}
\end{eqnarray}
where $f(k^2,p^2)$ is the form factor, which can, in principle, derived from lattice simulation. I use here $\displaystyle f(k^2,p^2)=\frac{e^{-m^2(k^2+p^2)/2}}{(2\pi)^3}$ for simplicity and parametrize the running coupling $\alpha(q)$ as\cite{SF11b}
\[
\alpha_s(q)=\frac{\gamma n(q)}{\log (\frac{q^2+m_g(q)^2}{\Lambda^2})}
\]
where $\gamma=\frac{12}{33-8}$, $n(q)=\pi(1+[\frac{\gamma}{\log(m^2/\Lambda^2)(1+q/\Lambda)-\gamma}+(bq)^c]^{-1})$, $m_g(q)=\frac{m}{1+(aq)^d}$, $m=1.024$GeV, $a=3.008$GeV$^{-1}$, $d=0.840$, $b=1.425$GeV$^{-1}$,$c=0.908$
and $\lambda=0.349$GeV.
  
The self-energy $\Pi^b_{22}$ derived from $J_b(k_x,p_z)$ is similar.  A numerical calculation in the above model, yields the self-energy $M/\alpha_s\sim 40$ eV.

\section{Decay processes}
In this section, phenomenological baryon decay, meson decay and neutrino mixing are investigated. 
\subsection{Baryon decay suppression}
In the MSSM, there are seven types of left-chiral superfields ${\mathcal Q}_i$ ${\mathcal H}_{ui}$, ${\mathcal H}_{di}$, ${\mathcal U}_{i}$,${\mathcal D}_{i}$, ${\mathcal L}_i$ and ${\mathcal E}_i$ where
$i$ denote families and it runs from 1 to 3.
\[
{\mathcal Q}_i=\left(\begin{array}{c}u\\
                                                 d\end{array}\right), \quad \left(\begin{array}{c}c\\  
                                                                                                                  s\end{array}\right),
\quad \left(\begin{array}{c}t\\
                                     b\end{array}\right)
\]
\[
\bar{\mathcal U}_i=\bar u, \bar c,\bar t
\]
\[
 \bar{\mathcal D}_i=\bar d,\bar s, \bar b
\]
\[
{\mathcal L}_i=\left(\begin{array}{c}\nu_e\\
                                                 e\end{array}\right), \quad \left(\begin{array}{c}\nu_\mu\\  
                                                                                                                  \mu\end{array}\right)
\quad \left(\begin{array}{c}\nu_t\\
                                     t\end{array}\right)
\]
\[
\bar{\mathcal E}_i=\bar e, \bar \mu,\bar \tau
\]

The $e$ and $\bar e^\dagger$ are the chilarity eigenstate: 
\[
\left(\begin{array}{c}e\\
                    \bar e^\dagger\end{array}\right)=\left(\begin{array}{c}e_L\\
                                                                                               e_R\end{array}\right)
\]

Under the ordinary MSSM, invariant superpotential $f_c^{abc}{\mathcal U}_1^a{\mathcal D}_1^b\circ{\mathcal D}_1^c$ which appears through a gauge transformation causes proton decay, which is not observed in the nature.
In order to make the MSSM consistent with the nature, the baryon triality 
\[
Z_3^B=e^{2\pi i[B-2Y]/3}
\]
where $B$ is the baryon number, and $Y$ is the hypercharge, is introduced as the discrete symmetry that should be preserved.

The transformation of the leptonic spinor field to vector field and vice versa in the octonion is different from that of the MSSM.

When a triality preservng product ${^tG_{123}}G_{12}{^tG_{13}}$ operates on $qq{^t\bar q}$,
\[
qq{^t\bar q}=\left(\begin{array}{c}C\\
                                             D\end{array}\right) \left(\begin{array}{c} A\\
                                                                                                  B\end{array}\right)
\left(\begin{array}{c}C\\
                             D\end{array}\right)\to 
\left(\begin{array}{c}E'(E)\\
                          E(E')\end{array}\right) \left(\begin{array}{c}E'(E) \\
                                                                                  E(E') \end{array}\right)
\left(\begin{array}{c}E'\\
                          E\end{array}\right)
\]
occurs. Here, $E(E')$ and $E'(E)$ can be identified as leptons on a light cone different from the original; i.e. the 4th component of one vector is twisted, in other words, $\xi_{14},\xi_{24},\xi_{34}$ are transformed to $-x_1',-x_2',-x_3'$ but $\xi_0$ is transformed to $x_4$. 
The two leptons and one vector particle production will be suppressed.

Similarly, when the transformation $G_{132}{^tG}_{13}G_{12}$ operates on $qq{^t\bar q}$, we find the proton decay into three vectors two of which are on a different light cone from the original
\[
qq{^t\bar q}=\left(\begin{array}{c}C\\
                                             D\end{array}\right) \left(\begin{array}{c} C\\
                                                                                                  D\end{array}\right)
\left(\begin{array}{c}A\\
                             B\end{array}\right)\to \left(\begin{array}{c}E'(E)\\
                                             E(E')\end{array}\right) \left(\begin{array}{c}E' \\
                                                                                                 E \end{array}\right)
\left(\begin{array}{c}E'(E)\\
                          E(E')\end{array}\right)
\]
will be suppressed.

In both cases, since intrinsic parity of the initial state is +, relative wave function of vector particles is required to be odd, and these processes are hindered.

When an triality changing transformation, e.g. $G_{23}G_{13}G_{123}$ operates on the three quark state $qq{^t\bar q}$ 
\[
qq{^t\bar q}=\left(\begin{array}{c}A\\
                                             B\end{array}\right) \left(\begin{array}{c} A\\
                                                                                                  B\end{array}\right)
\left(\begin{array}{c}C\\
                             D\end{array}\right)\to \left(\begin{array}{c}C\\
                                             D\end{array}\right) \left(\begin{array}{c}A(B) \\
                                                                                                 B(A) \end{array}\right)
\left(\begin{array}{c}A(B)\\
                             B(A)\end{array}\right),
\]
appears. When ${^t(}C,D)$ can be identified as $e^+$ and ${^t(}A(B),B(A))$  and ${^t(}A(B),B(A))$ make a meson $M_X$, one could interpret this process as proton decay into $e^+ M_X$\cite{Susskind05}. 

Thus, the stability of a baryon is assured when changing its triality sector is suppressed.  The transformation to another triality sector than before will not occur through normal electromagnetic processes.

\subsection{Meson decay}
The color neutral $q\bar q$ mesonic state consists of
$q^\alpha_a \bar q^{\bar \alpha}_{\bar a}$ and under the transformation $G_{23}{^tG}_{23}$, the transformation
\[
\left(\begin{array}{c}A\\
                            B\end{array}\right)\left(\begin{array}{c} C\\
                                       D\end{array}\right)\to\left(\begin{array}{c}C\\
                                                                                                  D\end{array}\right)
\left(\begin{array}{c}A\\
                             B\end{array}\right)
\]
occurs.  Under $G_{13}{^tG}_{13}$ 
\[
\left(\begin{array}{c}A\\
                            B\end{array}\right)\left(\begin{array}{c} C\\
                                       D\end{array}\right)\to\left(\begin{array}{c}A(B)\\
                                                                                                          B(A)\end{array}\right)
\left(\begin{array}{c}C(D)\\
                           D(C)\end{array}\right)
\]
In this case, twists in the 4th component of the spinor occur. One may identify ${^t(}A,B)$ as $u-$ quark and 
${^t(}C,D)$ as $\bar d$ and ${^t(}A(B),B(A))$ as $\mu$ and ${^t(}C(D),D(C))$ as $\bar \nu_{\mu}$, or
weak decay  $\pi^+\to \mu^+\nu_\mu$, and the overlap of the initial and the final state becomes weak.

 It means that the meson state of
$\left(\begin{array}{c}A\\
                              B\end{array}\right)\left(\begin{array}{c}C\\
                                                                                     D\end{array}\right)$ is electromagnetically stable, but it could decay via weak interaction.

The lagrangian for the $B$ meson decay in MSSM\cite{BK99} consists of direct coupling of the quark current and the lepton current, and via Higgs particle exchange.
\[
{\mathcal L}=-V_{qb}\frac{4G_F}{\sqrt 2}(\bar q\gamma^\mu\frac{1-\gamma_5}{2}b)(\bar l_k\gamma_\mu \frac{1-\gamma_5}{2}\nu_k)-R_l(\bar q\frac{1+\gamma_5}{2}b)(\bar l_k\frac{1+\gamma_5}{2}\nu_k)
\]
where $R_l=r^2 m_{l_k} m_b^Y$ and $r=\frac{\tan\beta}{m_H}$. If right-handed leptons, which could present after the Big bang is at present absent, one could choose $r=0$.

In the direct coupling part, under the transformation $G_{12}{^tG}_{12}$, the configuration  $b \bar d$ expressed by the basis ${^t(}A,B){^t(}C,D)$ will be transformed to ${^t(}A(B),B(A)) {^t(}C(D),D(C))$, which is interpleted as $\l\nu_{\l}$. The presence of twist in the 4th component will suppress the transition.
\subsection{Neutrino mixing}

In the standard model, the Cabibbo-Kobayashi-Maskawa quark mixing matrix 
\[
U_{CKM}=V_L^{d\dagger} V_L^u
\]
is neary diagonal, but the leptonic $3\times 3$ unitary mixing PMNS (Pontecorvo-Maki-Nakagawa-Sakata) matrix\cite{Ponte57,MNS62,Ponte68} 
\[
U_{PMNS}=V_L^{\l\dagger} U_{TBM} W
\]
where $U_{TBM}$ is the lepton mixing matrix, and $W$ is the neutrino mixing matrix,  has large off diagonal components.  In  $V_L^{\l}$ the mixing between $\nu_e$ and $\nu_{\tau}$ is small,  

Experimentally, $\nu_\mu\to\nu_e$ oscillation is large in the LSND collabolation\cite{LSND01} where the 800MeV  proton beam is used, but no evidence in the MiniBooNE collaboration\cite{MB09} where the 8GeV proton beam is used. 
Theoretically, contribution of $\nu_\tau$ in $\nu_e, \nu_\mu$ coupling was proposed in \cite{BPW95},  and  the LSND collaboration showed possible contribution of the sterlile neutrino.  MiniBooNe collaboation showed that the 3 neutrino + 2 sterile neutrino coupling model reproduces the LSND and the MiniBooNe data\cite{MB04,MB07,MB09}.
 The recent T2K experiment\cite{T2K11} suggests that the coupling of $\nu_e$ and $\nu_\tau$ is relatively large. 

The origin of sterile neutrino is not well known. 
As \cite{MPS08}, we consider the unitary transformation choosing the basis $\alpha=e,\mu,\tau,s,p$ where sterile leptons $s,p$ are assumed. 
The difference of mass squared of the sterile neutrino $\nu_4/\nu_5$ and that of $\nu_1$ is defined as $\Delta m_{41/51}^2=m_{4/5}^2-m_{\nu_1}^2$ is expected to be of the order of 1eV$^2$.  Since the mass is not large as compared to $\nu_e$ or $\nu_\mu$, I assume they are neutrinos in triality sectors other than that of $e$ or $\mu$, not necessarily right-handed as \cite{MPS08}.

 The physical neutrino states are expressed as $\nu_\alpha=U_{\alpha i}\nu_i$\cite{MPS08}, where
\[
U_{\alpha i}=\left(\begin{array}{ccccc}U_{e1}& U_{e2}& U_{e3}&U_{e4}&U_{e5}\\
                                                U_{\mu 1}& U_{\mu 2}& U_{\mu 3}&U_{\mu 4}&U_{\mu 5}\\
                                                U_{\tau 1}& U_{\tau 2}& U_{\tau 3}&0 &0\\
                                                0& 0 & 0& 1&0\\
                                                0& 0&0 &0&1\end{array}\right)=
\left(\begin{array}{ccccc}0.81& 0.55& 0 &U_{e4}&U_{e5}\\
                                                -0.51& 0.51& 0.70&U_{\mu 4}&U_{\mu 5}\\
                                                0.28& -0.67& 0.70&0 &0\\
                                                0& 0 & 0& 1&0\\
                                                0& 0&0 &0&1\end{array}\right).
\]

They showed that in high energy
\[
P_{\nu_\mu\to \nu_e}\propto \frac{1}{E^2}[(|U_{e4}U_{\mu 4}|\Delta m_{41}^2)^2+
   (|U_{e5}U_{\mu 5}|\Delta m_{51}^2)^2+(|U_{e4}U_{\mu 4}|\Delta m_{41}^2)(|U_{e 5}U_{\mu 5}|\Delta m_{51}^2)
\cos\delta
\]
can become small, when $|U_{e4}U_{\mu 4}|\Delta m_{41}^2=|U_{e 5}U_{\mu 5}|\Delta m_{51}^2$ and $\delta\sim \pi$\cite{GGM07}. In the fit of \cite{MPS08} $\Delta m_{21}^2=8\times 10^{-5}$eV$^2$, $\Delta m_{31}^2=2.5$eV$^2$ are used. 
When the coupling between the physical states 1 and states in the different triality sector is parametrized as $\sin\phi_{41}$ or $\sin\phi_{51}$,  the decay probability becomes,
\begin{eqnarray}
P_{\nu_\mu\to \nu_e}&=&4|U_{e4}|^2|U_{\mu 4}|^2\sin^2 \phi_{41}+4|U_{e5}|^2|U_{\mu 5}|^2\sin^2 \phi_{51}\nonumber\\
&&+8|U_{e4}U_{\mu 4}U_{e5}U_{\mu 5}|\sin \phi_{41}\sin\phi_{51}\cos(\phi_{54}-\delta)
\end{eqnarray}
where $\displaystyle \phi_{ij}=\frac{\Delta m_{ij}^2 L}{4E}$ and $\delta=arg(U_{e4}^*U_{\mu 4}U_{e5}U_{\mu 5}^*)$.

The neutrino mixing angles are estimated to be $\theta_{23}\sim 45^\circ$, $\theta_{12}\sim 34^\circ$ and
recently measured Daya Bay data fixes $\theta_{13}\sim 9^\circ$\cite{DB12}.  
The fitting $U_{e3}=0$ of \cite{MPS08} will be modified to $U_{e3}=\sin\theta_{13}e^{-i\delta}\sim 0.156$, and
slight changes in other channels are expected.  
 When the masses of the two sterile neutrino are 0.68 eV and 0.95 eV, respectively, parameters $U_{e4}=-0.037$, $U_{\mu 4}=0.0077$, $U_{e5}=0.13$, $U_{\mu 5}=0.19$ were assigned to fit the data of  MiniBooNe\cite{MB04,MB07}.

A standard method of explaining non-zero mass of a neutrino is incorporation of right-handed heavy leptons and the see-saw mass generation mechanism\cite{Zuber12,MP98}. This model was extended recently to $6\times 6$ flavor mixing model\cite{Xing11, Xing12}. However, similar three flavor + three sterile neutrino model was discussed in \cite{MS07} and no significant improvement to 3 +2  flavor model was found. Whether right handed heavy leptons are necessary to understand mass hierarchy of neutrino is a problem.  

The CP violation and the energy dependence of $\nu_\mu\to \nu_e$ oscillation  in the 3+2 flavor model is discussed in \cite{GGM07}. More detailed interpretation of the two sterile neutrino as neutrinos in two different triality sectors is under investigation.

\section{The electromagnetic current and the neutral current}
 The neutrino interacts with quarks via the weak neutral current and the weak charged current.
The Slavnov identity for the electro-magnetic interaction of quarks suggests that the interaction is transversal. In low energy, the photon and the gluon could mix with each other and makes a color-flavor locked state. In high energy, massive Z boson instead of a photon could be exchanged and although the time component is canceled by non-physical ghost exchange, etc., longitudinal polarization as well as transverse polarization contributes in the propagation of the electromagnetic fields.

When a particle belongs to a definite triality sector, it will be selected by electro-magnetic probes of the same triality sector. The electromagnetic current\cite{PS95}
\begin{equation}
J_{EM}^\mu=\bar e\gamma^\mu(-1)e+\bar u\gamma^\mu(\frac{2}{3})u+\bar d\gamma^\mu(-\frac{1}{3})d
\end{equation}
selects one triality sector.  The neutral current in the standard model is expressed as
\begin{eqnarray}
J_Z^\mu=\frac{1}{\cos\theta_W}&[&\bar \nu_L\gamma^\mu(\frac{1}{2})\nu_L+\bar e_L\gamma^\mu(-\frac{1}{2}+\sin^2\theta_W)e_L+\bar e_R\gamma^\mu(\sin^2\theta_W)e_R\nonumber\\
&+&\bar u_L\gamma^\mu(\frac{1}{2}-\frac{2}{3}\sin^2\theta_W)u_L+\bar u_R\gamma^\mu(-\frac{2}{3}\sin^2\theta_W)u_R\nonumber\\
&+&\bar d_L\gamma^\mu(-\frac{1}{2}+\frac{1}{3}\sin^2\theta_W)d_L+\bar d_R\gamma^\mu(\frac{1}{3}\sin^2\theta_W)d_R ]
\end{eqnarray}
\begin{equation}
J_W^{\mu+}=\frac{1}{\sqrt 2}(\bar \nu_L\gamma^\mu e_L+\bar u_L\gamma^\mu d_L)
\end{equation}
\begin{equation}
J_W^{\mu-}=\frac{1}{\sqrt 2}(\bar e_L\gamma^\mu\nu_L+\bar d_L\gamma^\mu u_L)
\end{equation}
where $\theta_W$ is the Weinberg angle.

When the neutrino couples with a current in a different triality sector, the mixing matrix would become almost flavor independent. 

 Absence of triality symmetry in Majorana neutrinos, allows transformations among three triality sectors of currents that couple to neutrino, and large off diagonal component in neutrino mixing can be understood\cite{SF11a}.

The hypothesis that the dark matter are baryons in different triality sectors is consistent qualitatively  with the recent astronomical observation\cite{PABMT10} that the actual dark matter density in the solar neighborhood is on average 21\% larger than inferred from most dynamical measurements.

\section{The dark matter and the velocity of neutrino}
Around our solar system, dark matter density is higher than the spherical distribution\cite{PABMT10}. 
The matter may be baryons in different triality sectors, which are insensitive to the electromagnetic probes on the earth.
In our universe 70\% is dark energy, 25\% is dark matter and 5\% is the normal matter \cite{Sanders10}.  

Dark matter 25\% contents= 10\% normal matter in different triality sector+15\% neutrinos and other particles, whose $SU(2)_L$ leptonic partner belongs to different triality sectors than that of the electromagnetic detector.   

After 380000 years after the big bang, neutrino was thought to constitute about 10\% of the universe. This percentage does not seem to change much although dark energy increased very much after the epoch of 380000 years.

At the moment of the explosion of the Supernova 1987A,  neutrino arrived on the earth 4 hours earlier than the light arrived.
In September  2011, CERN OPERA experiment group announced that the $\nu_\mu$ traveled from CERN to the detector at Gran Sasso 60 ns faster than the light might have traveled\cite{OPERA11}. However, in March 2012, ICALUS and CERN collaboration presented a result of experiment using tightly bunched beam structure, which showed almost the same time of flight of a neutrino and a photon\cite{ICARUS12}, and at the moment OPERA experiment and ICALUS experiment are consistent.  

 A possible mechanism of extra luminous velocity is that the $\nu_\mu$ represented by the left-handed Weyl spinor $\chi_{\nu_\mu}$ makes an oscillation to $\nu_e$ but in the triality sector different from the original and make another oscillation to return to $\nu_\mu$: 
\[
\Psi_M=\left(\begin{array}{c} i\sigma^2\chi_{\nu_\mu}^{\dagger T}\\
                                                         \chi_{\nu_\mu}\end{array}\right)=
\left(\begin{array}{c}A_{\nu_\mu}\\                                                                                                                                                                      B_{\nu_\mu}\end{array}\right)\to \left(\begin{array}{c}A(B)_{\tilde\nu_e}\\                                                                                                                                                                      B(A)_{\tilde\nu_e}\end{array}\right)\to \left(\begin{array}{c}A_{\nu_\mu}\\                                                                                                                                                                      B_{\nu_\mu}\end{array}\right)
\]
The velocity of $\nu_\mu$ is $v=\frac{p}{E_{\nu_\mu}}=\frac{p}{\sqrt{p^2+m_{\nu_\mu}^2}}<1=c$. If the 4th
component is changed to $E_{\tilde \nu_e}=\sqrt{\tilde p^2+m_{\nu_e}^2}$ and $m_{\nu_e}<m_{\nu_\mu}$,  $v=\frac{p}{E_{\tilde\nu_e}}$ detected in a certain period during the path could become larger than $1=c$.
Possibility of CPT violation in the neutrino oscillation process is discussed in \cite{MB11}.

 It is interesting to check whether the interchange of the light cone with the 4th component $\xi_0$ to $\xi_{1234}$ really occurs via successive $G_{13}$ and $G_{13}^{-1}$ transformation.

\section{Conclusion and discussion}
I showed that the super symmetry associated with the octonion has the property different from that of MSSM and that the invariant potential does not contain the $f^{abc}{\mathcal U}_1^a{\mathcal U}_1^b  \circ{\mathcal D}_1^c$ term.  
The triality preserving transformation $G_{123}{^tG_{13}}G_{12}$ allows the $qq{^t\bar q}$ decay to $E(E')E E'$. However, since $E(E')$ is on the light-cone different from $E$ or $E'$, the process may not be detected on the earth, and they may appear as dark energy, which constitutes about 70\% of the total energy of the universe. 
Arbitrary operation of ${^tG}_{ijk}G_{ijk}$ or ${^tG_{ij}} G_{ij}$ on the baryons or mesons that consist of quarks will produce quarks in different light cones $A(B)$ etc., which may be assigned as the dark matter.

I showed that the neutral current $J_Z^\mu$ contains the coupling to neutrino and quarks. In the coupling to neutrinos or to a Majorana particle, the triality selection rule would not apply, but the coupling to quarks the selection rule works. That makes the difference in the CKM quark mixing and the neutrino mixing.

I showed that the three flavors + two sterile neutrinos model could be modified to three flavors + neutrinos in two different triality sectors could become a good candidate for explaining the LSND\cite{LSND01} and MiniBooNe\cite{MB09}.
The validity of our model could be checked by the neutrino oscillation and/or the detailed comparison of the neutral current contribution and the charged current contribution. 

Although it is hard to estimate quantitatively, the transverse gluon could have mass of a few tens of eV and the longitudinal gluon has a momentum dependent mass in matter. I estimated the mass of the transverse gluon using simple gaussian wave function of the quark. It may be interesting to use the hadronic form factor of ADS/QCD type models\cite{deTB12}.

In neutron star, the physics of the nuclear matter state and the CFL state are discussed in \cite{ARRW01}. The difference of the velocity of light and neutrino in the supernova 1987A is expected to be due to the interaction between matter in the supernova and the light.  Whether the velocity of the accelerator neutrino $\nu_\mu$ of about 3GeV remains as before after oscillation $\nu_\mu$ to $\nu_e$ could be an interesting problem, since the discrepancy of the OPERA\cite{OPERA11} and ICARUS\cite{ICARUS12} may be interpreted as contamination of $\nu_e$ in $\nu_\mu$ of OPERA wider range incident beam.

When quarks are insensitive to the triality, it will appear as three times larger flavor degrees of freedom in the lattice simulation.  In the domain wall fermion lattice simulation, we selected one triality sector contribution via a rotation in the 5th dimension\cite{SF09}.


\newpage
\section*{Acknowledgements}

I received helpful information on neutrino physics at the Sapporo Winter School 2012 held at Hokkaido university in March 8-10 and at the KEK Theory Symposium 2012 held at KEK in March 5-7. I thank the organizer of the two work shops. I thank Stan Brodsky for helpful comments.  The support of numerical simulation using computers in KEK YITP at Kyoto university and in Tsukuba University are thankfully acknowledged.









\end{document}